\documentclass[twocolumn,showpacs,preprintnumbers,amsmath,amssymb]{revtex4}

\usepackage{graphicx}
\usepackage{dcolumn}
\usepackage{bm}

\begin{document}

\preprint{}

\title{Antiferromagnetic Vortex Core of Tl$_2$Ba$_2$CuO$_{6+\delta}$ Studied by NMR}

\author{K.~Kakuyanagi$^1$, K.~Kumagai$^1$, Y.~Matsuda$^2$, and M.~Hasegawa$^3$}

\affiliation{$^1$Division of Physics, Graduate School of Science, Hokkaido
University,  Sapporo 060-0810, Japan}%
\affiliation{$^{2}$Institute for Solid State Physics, University of Tokyo,
Kashiwanoha 5-1-5, Kashiwa, Chiba 277-8581, Japan}%
\affiliation{$^{3}$Institute of Material Research, Tohoku University, Katahira,
Sendai, 980-8577, Japan}%


\begin{abstract}

Spatially-resolved NMR is used to probe the magnetism in and around vortex cores of nearly optimally-doped Tl$_2$Ba$_2$CuO$_{6+\delta}$ ($T_c$=85~K).  The NMR relaxation rate $T_1^{-1}$ at $^{205}$Tl site provides a direct evidence that the AF spin correlation is significantly enhanced in the vortex core region.   In the core region Cu spins show a local AF ordering  with moments  parallel to the layers at $T_N$=20~K.   Above $T_N$ the core region is in the paramagnetic state which is a reminiscence of the state above the pseudogap temperature ($T^*\simeq$120~K), indicating that the pseudogap disappears within  cores.

\end{abstract}
\maketitle
	  In high-$T_c$ cuprates (HTC)  the superconductivity with $d$-wave symmetry appears when carriers are doped into the antiferromagnetic (AF) Mott insulators.   It is well established that the strong AF fluctuation plays a crucial role in determining many physical properties.    Therefore the relation between superconductivity and magnetism has been a central issue in the physics of HTC.   Especially, how the antiferromagnetism emerges when the $d$-wave superconducting order parameter is suppressed is a fundamental problem in the superconducting state \cite{so5,sachdev}.  In this respect, the microscopic structure of  vortex core, which is a local normal region created by destroying the superconductivity by magnetic field, turns out to be a very interesting subject.

	Within the framework of the semiclassical approximation,  in which the electron correlation effects are  ignored,  vortex cores in $d$-wave superconductors are in the normal metallic state which is same as the state above $T_c$, similar to $s$-wave superconductors \cite{semiclassic}.    However, recent high resolution STM experiments have revealed many unexpected properties in the spectrum of  vortex cores, which are fundamentally different from these semiclassical $d$-wave vortex cores \cite{stm}.   For instance, a checkerboard halo of the local density of states (LDOS) around the core has been reported in Bi$_2$Sr$_2$CaCu$_2$O$_{8+\delta}$ \cite{hoffman}.   A new class of theories has pointed out that the strong electron correlation effects change the vortex core structure dramatically.  For example, possible competing orders, such as AF \cite{AF}, staggered flux \cite{stag}, and stripe \cite{sachdev,stripe} orderings in and around cores have been discussed.   Therefore it is crucial for gaining an understanding of  the vortex state of HTC to clarify how the AF correlation and pseudogap phenomena, which characterize the magnetic excitation in the normal state, appear in and around vortex cores.

	Despite extensive studies, little is known about the microscopic electronic structure of the vortices, especially concerning the magnetism. The main reason for this is that  STM experiments do not directly reflect the magnetism.  Neutron scattering experiments on La$_{2-x}$Sr$_x$CuO$_4$ have reported that an applied magnetic field enhances the AF correlation in the superconducting state \cite{lake}.  However, the relation between the observed AF ordering and the magnetism within vortex cores is not clear, because the neutron experiments lack spatial resolution.  Recent $\mu$SR experiments on underdoped YBa$_2$Cu$_3$O$_{6.5}$ have reported the presence of static magnetism in vortex core region \cite{miller}, but the detailed nature of  this magnetism is still not clear. 
	
	Recent experimental \cite{curro,mitrovic,kakuyanagi} and theoretical \cite{takigawa} NMR studies have established that the frequency dependence of spin-lattice relaxation rate $T_1^{-1}$ in the vortex state serves as a probe for the low energy excitation spectrum which can resolve {\it different spatial regions of the vortex lattice}.    Unfortunately, up to now,  all of these spatially-resolved NMR measurements have been carried out at the planar $^{17}$O sites \cite{curro,mitrovic,kakuyanagi}, at which the AF fluctuations are filtered due to the location of O-atoms in the middle of neighboring Cu atoms with antiparallel spins \cite{masashi}. 
	
	 In this Letter we provide local information on the AF correlation in the different regions of the vortex lattice extending the measurements to the vortex core region, by performing a spatially resolved NMR imaging experiments on $^{205}$Tl-nuclei in nearly optimally-doped Tl$_2$Ba$_2$CuO$_{6+\delta}$. This attempt is particularly suitable for the above purpose because $T_1^{-1}$ at the Tl-site, $^{205}T_1^{-1}$, can monitor AF fluctuations sensitively. Quite generally, $1/T_1$ is expressed in terms of the dynamical susceptibility as $\frac{1}{T_1}=\frac{\gamma_nk_BT}{2\mu_B^2}\sum_{q}|A_q|^2\frac{Im\chi(q,\omega_
0)}{\omega_0}$, where $\gamma_n$ is the nuclear gyromagnetic ratio, $A_q$ is the hyperfine
coupling between nuclear and electronic spins, and $\omega_0$ is the Larmor
frequency.  Because Tl atoms are located just above the Cu atoms and there exist large transferred hyperfine interactions between Tl nuclei  and Cu spin moments through apical oxygen, Tl sees the full wavelength spectrum of Cu magnetic spin fluctuation; $^{205}T_1$ is dominated by $\chi$({\boldmath $q$}) at {\boldmath $q$} $=(\pi,\pi)$, {\it i.e.} AF fluctuations. This should be contrasted to the O-sites at which $\chi$({\boldmath $q$}) is dominated by uniform fluctuations at {\boldmath $q$}$=(0,0)$.  

\begin{figure}
\includegraphics [scale=0.43,angle=0] {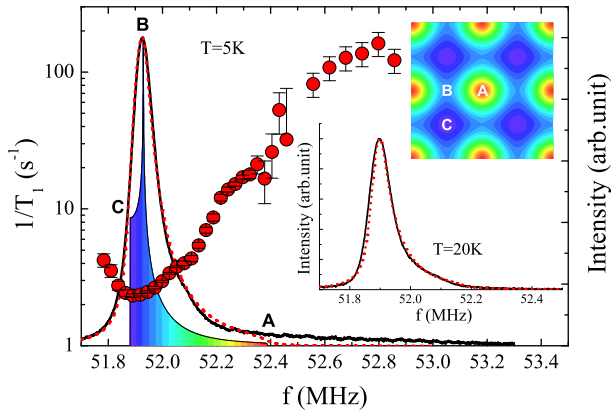}
\caption{Main panel: $^{205}$Tl-NMR spectrum (solid line) at 5~K.  The intensity is plotted in a linear scale. The thin solid line depict the histogram at particular local fields obtained from Eq.(1).   In the calculation we used $\xi_{ab}$=18\AA~and $\lambda_{ab}=1700~\AA$. The red dotted line represents the simulation spectrum convoluted with Lorentzian broadening function.  The red filled circles show the frequency dependence of $T_1^{-1}$  at the Tl site.      For details, see the text.  The lower inset:  $^{205}$Tl-NMR spectrum (solid line)and the simulation spectrum (red dotted line) at 20~K.   In the calculation we used $\xi_{ab}$=18.5\AA~and $\lambda_{ab}=2000~\AA$.  The upper inset: the image of the field distribution in the vortex square lattice; center of vortex core (A), saddle point (B) and center of vortex lattice (C). }
\end{figure}

	 NMR measurements were carried out on the $c$-axis oriented polycrystalline powder of high quality Tl$_2$Ba$_2$CuO$_{6+\delta}$ ($T_c$=85~K) in the external field  ($H_0$=2.1~T) along the $c$-axis. The $^{205}$Tl spin echo signals were obtained by a pulse NMR spectrometer.  The spectra was obtained by convolution of the respective Fourier-transform-spectra of the spin echo signals measured with an increment of 50~kHz. A very sharp spectrum ($\sim$ 50~kHz) above $T_c$ becomes broad below $T_c$ due to the development of vortices.    The solid lines in Fig.~1 and lower inset depict the NMR spectra at 5~K and 20~K, respectively.   A clear asymmetric pattern of the NMR spectrum, which originates from the local field distribution associated with the vortex lattice  is observed below the vortex lattice melting temperature ($\sim$60~K at $H_0$) \cite{FCC}.   The local field profile in the vortex state is given by approximating $H_{loc}(\mbox{\boldmath$r$})$ with
the London result,
\begin{equation}
H_{loc}(\mbox{\boldmath$r$})=H_0\sum_{G}\exp{(-i\mbox{\boldmath$G\cdot
r$})}~\frac{\exp({-\xi_{ab}^2\mbox{\boldmath$G$}^2/2})}
{1+\mbox{\boldmath$G$}^2\lambda_{ab}^2},
\end{equation}
where $\mbox{\boldmath$G$}$ is a reciprocal vector of the vortex lattice, $\mid\mbox{\boldmath$r$}\mid$ the distance from the center of the core, $\xi_{ab}$  the in-plane coherence length, and $\lambda_{ab}$  the in-plane penetration length.   The thin solid lines  in Fig.~1 depicts  the histogram at a particular local field which is given by the local field distribution $f(H_{loc})=\int_{\Omega} \delta [H_{loc} (\mbox{\boldmath$r$})-H_{loc}]d^2\mbox{\boldmath$r$}$ in Eq.(1) assuming the square vortex lattice, where $\Omega$ is the magnetic unit cell.  The upper inset shows the image of the field distribution in the vortex lattice.  In the prefect vortex lattice and without  magnetism within cores which will be discussed later, the histogram shows the low and high frequency cutoffs at the center of the vortex lattice (C-point) and at the center of cores (A-point), respectively, and shows a peak at the field corresponding to the saddle point  (B-point).     This characteristic spectrum  (Redfield pattern) demonstrates that the NMR frequency depends on the position of the vortex lattice. We therefore can obtain the spatially-resolved information of the low energy excitation by analyzing the frequency distribution of the corresponding NMR spectrum.  

\begin{figure}
\includegraphics [scale=0.35,angle=0] {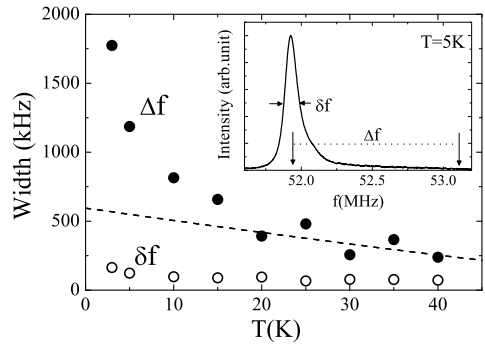}
\caption{The line widths of the $^{205}$Tl-NMR spectrum plotted as a function of $T$. $\Delta f$ (filled circles) indicates the line width determined by the frequency at which the intensity becomes 1\% of the peak. The dashed line represents the line width calculated from the simulation spectrum without taking into account the core magnetism.   Open circles represent $\delta f$ which is defined as the line width at the half intensity.  Inset shows the spectrum at 5~K and the definition of $\Delta f$ and $\delta f$.}
\end{figure}

	We first discuss the observed NMR spectra. The real spectrum broadens due to the imperfect orientation of the power and distortion of the vortex lattice.   The red dotted lines in Fig.~1 and the lower inset  represent the simulation spectra in which  the Lorentzian broadening function, $f(H_{loc})=\sigma/(4H_{loc}^2 + \sigma^2)$, is convoluted to Redfield pattern using $\sigma$=42~kHz.   The theoretical curve reproduces the data  well in the whole frequency range at $T$=20~K (inset).   On the other hand,  the spectrum at 5~K shows significant broadening at high frequency region (core region), while it can be well fitted below 52.1~MHz (main panel).    The filled circles in Fig.~2 display the line width at the high frequency tail, $\Delta f$, which is defined as a difference between the frequency at the peak intensity  and the frequency at which the intensity becomes 1\% of the peak.  For the comparison,  the line width calculated from the simulation spectra with Lorentzian broadening function is plotted by the dashed line.  At high temperatures $\Delta f$ agrees well with the calculation, while below 20~K it becomes much larger.  We also plot $\delta f$ which represents the line width at the half intensity (open circles).   The fact that $\delta f$ changes little below 20~K confirms that  the line broadening occurs only in the high frequency core region.   In $\mu$SR experiments,  the high frequency tail was attributed to the static magnetism around cores, which causes additional broadening \cite{miller}.   It should be noted that because of large transfer hyperfine coupling between Tl nuclei and Cu moment, the broadening of Tl-NMR spectrum associated with the static magnetism is more pronounced  than that of $\mu$SR spectrum.   Therefore the observed broadening below $\sim$20~K is naturally explained by  the appearance of static magnetism within cores below $\sim$20~K.   To obtain deep insight into this phenomena, the measurements of $T_1$ are crucial.  

\begin{figure}
\includegraphics [scale=0.39,angle=0] {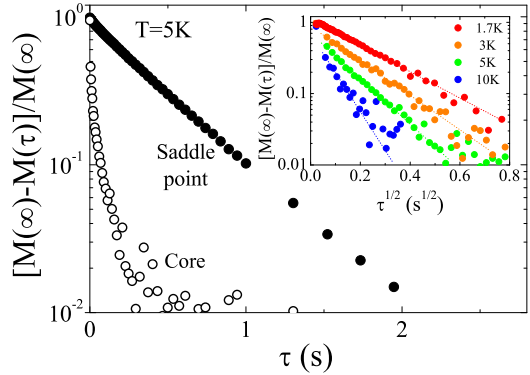}
\caption{Recovery curves of nuclear magnetization of $^{205}$Tl  as a function of time, $\tau$.   The filled and open circles represent the data at the saddle points and at vortex cores, respectively.  The inset shows the recovery curves at cores as a function of $\sqrt \tau$ at several temperatures.  }
\end{figure}

	For $^{205}$Tl with nuclear spin $I$=1/2, the recovery curve of the nuclear magnetization $M(t)$ fits well to a single exponential relation, $R(\tau)=(M(\infty)-M(\tau))/M(\infty)= \exp(-\tau/T_1)$ in the normal state. In the vortex state, on the other hand, the feature of the recovery curves is strongly position dependent as shown in Fig.~3.   The spin echo intensities are measured as a function of $\tau$ after saturation pulses. Then the nuclear magnetization recovery curves are obtained from each frequency component of the Fourier transform spectra.  We obtained the data set of the recovery intensity for each frequency point at the 28~kHz interval with a gaussian weight function of $\sigma$=10~kHz. There are two distinct features. First, the decay time at cores is much faster than at saddle points.  Second, while the recovery curves show the single exponential at the saddle point, they show a $\sqrt{\tau}$ dependence at the core region as shown in the inset of Fig.~3.  We will discuss this $\sqrt{\tau}$ dependence later.   In what follows, we defined $T_1$ as the time required for the nuclear magnetization to decay by a factor 1/$e$, in order to define $T_1$ uniquely for either decay curve.
		
	The red filled circles in Fig.~1 show the frequency dependence of $^{205}T_1^{-1}$. On scanning from  outside into cores, $^{205}T_1^{-1}$ increases rapidly after showing a minimum near saddle points.  The magnitude of $^{205}T_1^{-1}$ in the core region is almost two orders of magnitude larger than that near saddle point. This large enhancement of $^{205}T_1^{-1}$ is in striking contrast to $^{17}T_1^{-1}$ at $^{17}$O sites reported in YBa$_2$Cu$_3$O$_7$ \cite{curro,mitrovic} and  YBa$_2$Cu$_4$O$_8$ \cite{kakuyanagi}, in which the enhancement of $^{17}T_1^{-1}$ at the core region is 2-3 times at most  and has been attributed to LDOS produced by a Doppler shift of the QP energy spectrum by supercurrents around the vortices \cite{volovik}.  It should be noted that the LDOS effect is absent in $^{205}T_1^{-1}$, because there are no conduction electrons at $^{205}$Tl-site.  Therefore, the remarkable enhancement of $^{205}T_1^{-1}$ provides a direct evidence that {\it the AF correlation is strongly enhanced near the vortex core region} \cite{vib}. The decrease of $^{205}T_1^{-1}$ well outside the core when going from point C to B was also reported in $^{17}T_1^{-1}$ \cite{mitrovic,kakuyanagi}.  
		
	Figures 4 (a) and (b) depict the $T$-dependences of  $(^{205}T_1T)^{-1}$ and $^{205}T_1^{-1}$ within cores (filled circles) and at the frequency corresponding to saddle points (open circles), respectively.   In these figures, $^{205}T_1^{-1}$ within cores, $(^{205}T_1^{core})^{-1}$,  were determined with using integrated intensities over the high frequency region beyond the A point.  From high temperatures down to about 120~K,  $(^{205}T_1T)^{-1}$ obeys the Curie-Weiss law, $(^{205}T_1T)^{-1}\propto1/(T+\theta)$. The lowest $T$ at which this law holds is conveniently called the pseudogap temperature $T^*$.   Below $T^*$,  $(^{205}T_1T)^{-1}$ decreases rapidly without showing any anomaly associated with the superconducting transition at $T_c$, similar to other HTC \cite{masashi}.   

\begin{figure}
\includegraphics [scale=0.44,angle=0] {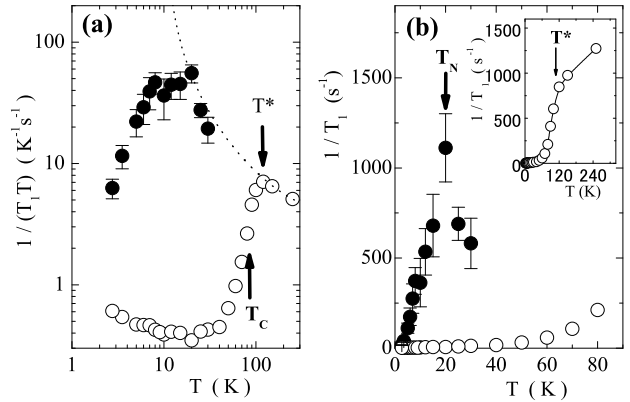}
\caption{ (a) $T$-dependence of $(^{205}T_1T)^{-1}$.  (b) $T$-dependence of $^{205}T_1^{-1}$ at low temperatures (main panel) and at high temperatures (inset) .  The filled and open circles represent the data at vortex cores and at  saddle points, respectively.   $T^*\simeq 120$~K is the pseudogap temperature. The dotted line represents the Curie-Weiss law which is determined above $T^*$. In (b), $T_N$ is the temperature at which $^{205}T_1^{-1}$ at the core exhibits a sharp peak.}
\end{figure}

	The $T$-dependence of $(^{205}T_1^{core})^{-1}$ contains some key features for understanding the core magnetism.   The first  important signature is that $1/^{205}T_1^{core}$ exhibits a sharp peak at  $T$=20~K, which we label as $T_N$ for future reference (Fig.~4(a)).    Below $T_N$ $(^{205}T_1^{core})^{-1}$ decreases rapidly with decreasing $T$.   There are two possible origins for this peak.  One is the reappearance of the pseudogap and the other is the occurrence of a local static AF (or SDW) ordering in the core region.   Generally $T_1^{-1}$ shows a sharp peak when the AF ordering occurs.  The fact that the sharp peak of $(^{205}T_1^{core})^{-1}$ is observed at $T_N$ while $1/T_1$ at the saddle points shows neither a peak nor broad maximum at $T$* as shown in the inset of Fig.~4(b) excludes the possibility of pseudogap, supporting the AF ordering.   Moreover, as shown in Fig.~2,  broadening of the NMR spectrum starts at $\sim$20~K, which coincides with $T_N$.  This fact gives an additional strong evidence on AF ordering.   We also point out that the appearance of the local AF ordering is also consistent with the $\sqrt{\tau}$ dependent nuclear magnetization decay curve shown in the inset of Fig.~3.  In fact, the $\sqrt{\tau}$ dependence has been observed when the microscopic imhomogeneous distribution of $T_1^{-1}$ due to strong magnetic scattering centers is present \cite{sil}.  On basis of these results, we are lead to conclude that the local AF ordering takes place in the core region at $T_N$=20~K; $T_N$ {\it corresponds to the N\'eel temperature within the core.} This AF ordering is  consistent with the prediction of recent theories based on the $t-J$ and SO(5)  models \cite{AF}.  The present results also should be distinguished from those of the neutron scattering experiments on La$_{2-x}$Sr$_x$CuO$_4$ \cite{lake}, in which the static SDW coexists with superconductivity even in zero field just below $T_c$.  In the present compound, on the other hand, we do not observe such a static SDW ordering and the vortex core region is in the paramagnetic state in a wide $T$-region between $T_c$ and $T_N$. 
	
	As discussed before, the broadening occurs only at high frequencies. This fact indicates that {\it the AF spins are oriented parallel to the CuO$_2$ layers}. This follows by observing that the broadening should occur at both high and low frequency sides if the AF ordering occurs perpendicular to the layers, because in this case the direction of the alternating transferred hyperfine fields are parallel and antiparallel to the applied field. Using the hyperfine coupling constant, $A_{hf}$= 65kOe/$\mu_B$, the magnetic moments induced within the core is estimated to be $\sim$0.1$\mu_B$ at $H_0$.  The detailed analysis will be published in elsewhere.  

	The second important signature for the core magnetism is that, as shown by the dotted line in Fig.~4(a),  $(^{205}T_1^{core}T)^{-1}$ above $T_N$ nearly lies on the Curie-Weiss law line extrapolated above $T^*$.  This fact indicates that the vortex core region appears to be {\it in the paramagnetic state which is a reminiscence of the state above $T^*$; the pseudogap is absent in the core region.}     This result seems to be inconsistent with the recent theories which predict  local orbital currents, in which the pseudogap phenomenon within cores is assumed \cite{AF}.

       Summarizing the salient features of spatially-resolved NMR results in the vortex lattice;(1) NMR spectrum near the core region broadens below $T$=20~K (Figs.~1 and 2).   (2) Upon approaching the vortex core, $(^{205}T_1)^{-1}$ is strongly enhanced (Fig.~1).  (3) Near the core region, the NMR recovery curves show the $\sqrt{\tau}$-dependence (Fig.~3).  (4) $(^{205}T_1^{core})^{-1}$ exhibits a sharp peak at $T$=20~K (Fig.~4),  All of these results provide direct evidence that in the vortex core region the AF spin correlation is extremely enhanced, and that the paramagnetic-AF ordering transition of the Cu spins takes place at $T_N=20$~K.   We also find the pseudogap disappears within the core.  The present results offer a new perspective on how the AF vortex core competes with the $d$-wave superconductivity.

	We acknowledge helpful discussions with M.~Franz, M.~Imada, J.~Kishine, K.~Machida, D.K.~Morr, M.~Ogata,   S.H.~Pan, M.~Takigawa, A.~Tanaka, Z.~Tesanovi\'c, and O.M.~Vyaselev.


\begin{thebibliography}{50}

\bibitem{so5} S.C.~Zhang, Science 275, 1089(1997)
\bibitem{sachdev}Y.~Zhang, E.~Demler, and S.~Sachdev,  Phys. Rev. B. {\bf 66}, 094501 (2001) and references therein.
\bibitem{semiclassic}N.~Schopohl, and K.~Maki, Phys. Rev. B {\bf 52}, 490 (1995), M.~Ichioka, {\it et al.} Phys. Rev. B {\bf 53}, 15316 (1996).
\bibitem{stm}I.~Maggio-Aprile, {\it et al.} Phys. Rev. Lett. {\bf 75}, 2754 (1995), S.H.~Pan,{\it et al.}  {\it ibid} {\bf 85}, 1536 (2000).
\bibitem{hoffman}J.E.~Hoffman {\it et al.}, Science {\bf 295}, 466 (2002).
\bibitem{AF}D.P.~Arovas, {\it et al.} Phys. Rev. Lett. {\bf 79}, 2871 (1997), J.H.~Han and D.H.~Lee, {\it ibid.} {\bf 85}, 1100 (2000), Jian-Xin~Zhu and C. S.~Ting, {\it ibid} {\bf  87}, 147002 (2001),  Jian-Xin~ Zhu, {\it et al.} {\it ibid.}  {\bf 89}, 067003 (2002), M.~Ogata, Int. J. Mod. Phys. B {\bf 13}, 3560 (1999), H.~Tsuchiura {\it et al.} cond-mat/0302030.
\bibitem{stag} J.~Kishine {\it et al.}, P.A.~Lee PA, X.G.~Wen, Phys. Rev. Lett. {\bf 86}, 5365 (2001), Q.H.~Wang {\it et al.}  {\it ibid.} {\bf 87}, 167004 (2001).
\bibitem{stripe} M.~Takigawa {\it et al.} Phys. Rev. Lett. {\bf 90}, 047001 (2003).
\bibitem{lake}B. Lake,{\it et al.} Science {\bf 291}, 1759 (2001), B. Lake, {\it et al.} Nature, {\bf 415}, 299 (2002), B. Khaykovich {\it et al.} Phys. Rev. B {\bf 66}, 014528 (2002).
\bibitem{miller}R.I.Miller {\it et al.} Phys. Rev. Lett. {\bf 88} 137002, (2002).
\bibitem{curro}N.J.~Curro, {\it et al.} Phys. Rev. B {\bf 62}, 3473 (2000).
\bibitem{mitrovic} V. F.~Mitrovic, {\it et al.}, Nature {\bf 413}, 501 (2001), V. F.~Mitrovic, {\it et al.} cond-mat/0202368
\bibitem{kakuyanagi} K. ~Kakuyanagi {\it et al.}, Phys. Rev. B {\bf 65}, 060503 (2002).
\bibitem{takigawa}M.~Takigawa {\it et al.}, Phys. Rev. Lett. {\bf 83}, 3057 (1999), R.~Wortis {\it et al.} Phys. Rev. B {\bf 61}, 12342 (2000), D.K.~Morr and R.~Wortis, Phys. Rev. B {\bf 61}, R882 (2000), and D.K.~Morr, Phys. Rev. B {\bf 63}, 214509 (2001), Y.~Chen{\it et al.} cond-mat/0302114 
\bibitem{masashi} V.~Barzykin and D.~Pines Phys. Rev. B {\bf 52}, 13585 (1995).
\bibitem{FCC} We stress  the importance of the measurements under field cooling condition in the vortex  lattice phase to observe the asymmetric field profile.  
\bibitem{volovik}G.E.~Volovik, JETP Lett. {\bf 58}, 469 (1993).
\bibitem{vib} We note that $(^{205}T_1^{core})^{-1}$ is two orders of magnitude larger than that expected solely from vortex vibration at all temperatures (see Fig.~2 in Ref.\cite{VV}).    
\bibitem{VV}L.N.~Bulaevskii {\it et al.} Phys. Rev. Lett. {\bf 71}, 1891 (1993).
\bibitem{sil} M.R. McHenry {\it et al.}, Phys.Rev. {\bf B5}, 2958 (1972).


\end{thebibliography}

%
\end{document}